\documentclass[showpacs,twocolumn,preprintnumbers,superscriptaddress,amsmath,amssymb,nofootinbib]{revtex4-1}

\usepackage[dvips]{graphicx}
\usepackage{color}
\usepackage{latexsym}
\usepackage{amsmath}
\usepackage{amssymb}
\usepackage{eufrak}
\usepackage{euscript}
\usepackage{pstricks}
\usepackage{graphics}
\usepackage{graphicx}
\usepackage{picture}
\def\[{\left\lbrack}
\def\]{\right\rbrack}

\def\({\left(}
\def\){\right)}

\newcommand{\bee}{\begin{equation}}
\newcommand{\eee}{\end{equation}}
\newcommand{\eaa}{\end{eqnarray}}
\newcommand{\baa}{\begin{eqnarray}}

\def\ni{\noindent}

\begin{document}

\title{\Large Quantum Gravitational Correction and MOND Theory in the Holographic Equipartition Scenario}

\author{Everton M. C. Abreu}\email{evertonabreu@ufrrj.br}
\affiliation{Grupo de F\' isica Te\'orica e Matem\'atica F\' isica, Departamento de F\'{i}sica, Universidade Federal Rural do Rio de Janeiro, 23890-971, Serop\'edica - RJ, Brasil}
\affiliation{Departamento de F\'{i}sica, Universidade Federal de Juiz de Fora, 36036-330, Juiz de Fora - MG, Brasil}
\author{Jorge Ananias Neto}\email{jorge@fisica.ufjf.br}
\author{Albert C. R. Mendes}\email{albert@fisica.ufjf.br}
\affiliation{Departamento de F\'{i}sica, Universidade Federal de Juiz de Fora, 36036-330, Juiz de Fora - MG, Brasil}

\date{\today}

\begin{abstract}
\noindent In this paper, by using Verlinde's formalism and a modified Padmanabhan's prescription, we have obtained the lowest order quantum correction to the gravitational acceleration and MOND-type theory by considering a nonzero difference between the number of bits of the holographic screen and number of bits of the holographic screen that satisfy the equipartition theorem. We will also carry out a phase transition and critical phenomena analysis in MOND-type theory where critical exponents are obtained.
\end{abstract}


\maketitle

\section{Introduction}

The formalism proposed by E. Verlinde\cite{Verlinde} obtains the gravitational acceleration  by using the holographic principle and the equipartition law of energy. His ideas relied on the fact that gravitation can be considered universal and independent of the details of spacetime microstructure.  Besides, he brought new concepts concerning holography since the holographic principle must unify matter, gravity and quantum mechanics. It is important to mention that similar ideas have also been given by Padmanabhan\cite{Pad2}. 

An important concept used by Verlinde is the notion of bits in which it can be  understood as the smallest unit of information in a holographic screen. Indeed, bits play an essential role in Verlinde formalism because when the total bits number is assumed to satisfy completely the equipartition law of energy, we have the well known formula of the classical gravitational acceleration. So, the aim of this work is to show that when we consider that the total number of bits no longer satisfies the equipartition theorem then, at first, we can obtain two important results which are both the quantum correction of the gravitational acceleration and MOND-type theory\cite{MOND1,MOND2,MOND3}.

In other words, we will obtain a constant parameter ($\alpha$) that is a measure of how much a difference between classical and quantum results concerning the number of bits is relevant.   We will see that this difference is not conserved at the quantum level.   In the second part of this work, the $\alpha$-parameter is a temperature function and a phase transition and critical phenomena analysis is provided.

This paper is organized in a way such that in section 2 we have depicted very briefly some Verlinde's entropic concepts.  In section 3 we have computed the $\alpha=N_S-N_E$ to show that $\alpha=0$ is not conserved at the quantum level.   In section 4 we have provided a brief review of MOND's main ideas.  In section 5 we have calculated the $\alpha$-parameter through MOND theory.   In section 6
we have carried out a phase transition and critical phenomena analysis on the holographic screen based on the computation of the $\alpha$-parameter.  In section 7 the conclusions are written.

\section{Some Entropic gravity ideas}

Before we begin to describe our proposal, let us review, in a short way, the Verlinde formalism. The model considers a spherical surface as being the holographic screen, with a particle of mass $M$ positioned in its center. The holographic screen can be imagined as a storage device for information. The number of bits, which is the smallest unit of information in the holographic screen, is assumed to be proportional to the  holographic screen
area $A$
\begin{eqnarray}
\label{bits}
N = \frac{A }{l_P^2},
\end{eqnarray}

\ni where $ A = 4 \pi r^2 $ and $l_P = \sqrt{\frac{G\hbar}{c^3}}$ is the Planck length and $l_P^2$ is the Planck area.  We can see clearly that the connection between $l_p$ and $\hbar$ characterizes $l_p$ as a quantum parameter.  It means that its introduction can be considered a semiclassical feature.
In Verlinde's framework one can suppose that the bits total energy on the screen is given by the equipartition law of energy

\begin{eqnarray}
\label{eq}
E = \frac{1}{2}\,N k_B T.
\end{eqnarray}
It is important to notice that the usual equipartition theorem in Eq. (\ref{eq}), can be derived from the usual Boltzmann-Gibbs thermostatistics. 
Let us consider that the energy of the particle inside the holographic screen is equally divided by all bits in such a manner that we can have the expression

\begin{eqnarray}
\label{meq}
M c^2 = \frac{1}{2}\,N k_B T.
\end{eqnarray}
With Eq. (\ref{bits}) and using the Unruh temperature equation  \cite{unruh} given by

\begin{eqnarray}
\label{un}
k_B T = \frac{1}{2\pi}\, \frac{\hbar a}{c},
\end{eqnarray}
we are  able to obtain the  (absolute) gravitational acceleration equation

\begin{eqnarray}
\label{acc}
a &=&  \frac{l_P^2 c^3}{\hbar} \, \frac{ M}{r^2}\nonumber\\ 
&=& G \, \frac{ M}{r^2}\,\,.
\end{eqnarray}
From Eq. (\ref{acc}) we can see that Newton's constant $G$ is just written in terms of the fundamental constants, $G=l_P^2 c^3/\hbar$.

\section{Quantum gravitational correction through Verlinde's concepts}

It is well known that the infrared behavior of quantum gravity is more interesting than the ultraviolet one.  However, the low energy propagation of massless particles leads us to quantum corrections for long distances.  In this way, the effective action may be expanded in a Taylor momentum expansion.  In the current literature, the standard calculation procedure is through Feynman rules.  In a different way, in
this section we will use the holographic and equipartition law ideas to compute the quantum correction to the Newtonian potential.

We begin our formalism  by rewriting expression  (\ref{eq}) in a similar form as the one from Padmanabhan\cite{Pad1}, except by the fact that we consider just the holographic screen, as
\begin{eqnarray}
\label{class}
\frac{A}{l_p^2}=\frac{E}{(1/2) k_BT}.
\end{eqnarray}

\ni Then, our proposal is to assume that there is a numerical difference between the total bits's number in the holographic screen and the bits' number that exist in the equipartition 
\begin{eqnarray}
\label{dif}
\frac{A}{l_p^2}-\frac{E}{(1/2)k_BT}=\alpha,
\end{eqnarray}
or
\begin{eqnarray}
\label{dif2}
N_S-N_E=\alpha,
\end{eqnarray}
where 

\begin{eqnarray}
\label{dif3}
N_S\equiv\frac{A}{l_p^2},
\end{eqnarray}
and

\begin{eqnarray}
\label{dif4}
N_E\equiv\frac{E}{(1/2)k_BT}.
\end{eqnarray}

\ni Hence, as we have said before, we can understand that the $\alpha$-parameter measures the difference between the quantum $N_S$ and the classical $N_E$ objects.   However, since $N_S$ and $N_E$ are two objects that provide one quantity, namely, the number of bits, naively one can say that $\alpha \rightarrow 0$.   But here we will make two assumptions.   The first one is that $\alpha << 1$ but not zero.  And the second one is that $N_S > N_E$, namely, $\alpha > 0$.  We believe that both assumptions are not very different from reality.  Since they are simple assumptions, any result that contradicts any one of them will be easily noticed.

Initially if we consider that the $\alpha$  parameter is a constant then (\ref{dif}) can be written as

\begin{eqnarray}
\label{conta1}
k_BT&=&\frac{2E\, l_p^2}{A-\alpha l_p^2}\nonumber\\\nonumber\\
&=&\frac{2E l_p^2}{A} \(1- \frac{\alpha l_p^2}{A}\)^{-1}.
\end{eqnarray}

\ni Performing a binomial expansion in (\ref{conta1}) we obtain that
\begin{eqnarray}
\label{conta12}
k_BT=\frac{2E l_p^2}{A} \(1+ \frac{\alpha l_p^2}{A}+...\),
\end{eqnarray}
where we have assumed that $ \frac{\alpha l_p^2}{A}<< 1$. Using the Unruh temperature formula, 
Eq.(\ref{un}), and $E=M c^2$ in  (\ref{conta12}), we can obtain a modified gravitational acceleration 

\begin{eqnarray}
\label{conta2}
a=\frac{G M}{r^2} \(1+ \frac{\alpha}{4\pi} \frac{l_p^2}{r^2}+...\).
\end{eqnarray}

\ni The second term in Eq.(\ref{conta2}) is the first order non-relativistic quantum correction to the gravitational acceleration.
There are several works that show different results for the coefficient $\alpha$ when general relativity is treated as an effective field theory. Among them we can mention works of Donoghue\cite{Don}, Akhundov et al\cite{Akh} and Bjerrum-Bohr et al\cite{BDH}.
As a simple potential cannot be considered an ideal relativistic concept, the general corrections would have the following expression

\begin{eqnarray}
\label{conta5}
V(r)=\frac{G m M}{r} \( 1+\beta \frac{l_p^2}{r^2}+... \),
\end{eqnarray}
where the $\beta$-parameter would rely on the exact definition of the potential [8].  It would be computed in the post-Newtonian expansion. The factor $l_p^2 /r^2$ is dimensionless and gives an expression parameter for the long distance quantum effects. The values of $\beta$ can be summarized in the table I\footnote{Extracted from S. Faller, Theorieseminar Universit\"{a}t K\"{o}ln-02.02.2009.}

\begin{table}[h]
\center
\caption{}
\begin{tabular}{lc}
\hline
Work & $\beta$ \\
\hline
Donoghue  & $\;-\frac{127}{30\pi^2}(\approx -0.43)$\\
Akhundov et al.  & $\;-\frac{107}{30\pi^2}(\approx -0.36)$ \\
Bjerrum-Bohr et al.  &\;\;\; $\frac{41}{10\pi}(\approx 1.31) $ \\
\hline
\end{tabular}
\end{table}

\vskip 1 cm
\noindent We will use that the gravitational acceleration is related to the potential through the equation 
$$ma=\left|\frac{dV}{dr}\right|,$$ then from (\ref{conta2}) and (\ref{conta5}) we can obtain a connection between $\beta$ and $\alpha$ given by

\begin{eqnarray}
\label{conta7}
\beta=\frac{\alpha}{12\pi}.
\end{eqnarray}

\ni Observing table I we can see that the first two negative results means that the number of bits that satisfy the equipartition theorem is greater than the total number of bits of the holographic screen. We consider that these results are not consistent with our proposal because we have assumed that $N_S\geq N_E$. Only the positive $\beta, (\beta=\frac{41}{10\pi})$ agrees with our model, i.e., $N_S\geq N_E$. We can imagine that $\alpha$ represents a small number of bits on the holographic screen. We have verified that if $\alpha=49$ then the value of $\beta$ obtained by Eq. (\ref{conta7}) approximately reproduces the value of $\beta$ given by Bjerrum-Bohr et al.  In fact, 49 bits is an extremely small value compared, as an example, with the number of bits on the holographic screen, Eq.(\ref{bits}), with radius 1 m that is $\approx 10^{70}$. Therefore, $\alpha=49$ can be interpreted as a fluctuation in the equality of the bits number of the holographic screen and the bits number that satisfies the equipartition theorem. A  small deviation of equality (\ref{class}) leads to a quantum correction in the gravitational acceleration. It is important to mention here that in our proposal
we are not using any usual procedure of a quantum field theory or effective field theory for the gravitational interaction. Only the holographic principle and the equipartition theorem which are the basis of the Verlinde formalism were used.

\label{MOND}

\section{MOND's review}

The MOND theory successfully explains the majority of the rotation curves of the galaxies. MOND theory reproduces the well known Tully-Fisher relation\cite{TF} and it can be also an alternative to dark matter. 

Basically, this theory is a modification of Newton's second law in which the force can be described as

\begin{eqnarray}
\label{mond1}
F=m \, \mu \(\frac{a}{a_0}\) a,
\end{eqnarray}
where $\mu(x)$ is a function with the following properties: $\mu(x)\approx 1$ for $x>>1, \mu(x)\approx x$ for $x<<1$ and $a_0$ is a constant. There are different interpolation functions for $\mu(x)$ \cite{prof1,prof2}. However, it is believed that the main implications caused by MOND theory do not depend on the specific form of these functions. Therefore, for simplicity, it is usual to assume that the variation of $\mu(x)$ between the asymptotic limits occurs abruptly  at $x=1$ or $a=a_0$. 

Let us consider \cite{JAN} that, below a critical temperature, the cooling of the holographic screen is not homogeneous. We choose that the fraction of bits with zero energy is given by the equation

\begin{eqnarray}
\label{mond2}
&\frac{N_0}{N}=1-\frac{T}{T_c},
\end{eqnarray}
where $N$ is the total number of bits given by the equation (\ref{bits}), $N_0$ is the number of bits with zero energy and $T_c$ is the critical temperature. For $T\geq T_c$ we have that $N_0=0$ and for $T\leq T_c$ the zero energy phenomenon for some bits starts to occur. Equation (\ref{mond2}) is a standard relation of critical phenomena and second order phase transitions theory. From (\ref{mond2}) the number of bits with energy different from zero for a given temperature $T<T_c$ is

\begin{eqnarray}
\label{mond3}
N-N_0=N\frac{T}{T_c}.
\end{eqnarray}
Considering that the energy of the particle inside the holographic screen is equally distributed over all bits with nonzero energy and using relation (\ref{mond2}) in the equipartition law of energy, we obtain that

\begin{eqnarray}
\label{mond4}
Mc^2=\frac{1}{2} N \frac{T}{T_c} k_B T.
\end{eqnarray}
Then, combining (\ref{bits}), (\ref{un}) and (\ref{mond4}), we are able to derive, for $T<T_c$, the MOND theory for Newton's law of gravitation\footnote{In references\cite{Ep1} and \cite{Ep2} the authors derive MOND's theory by considering that the bits of the holographic screen are fermionic excitations.}

\begin{eqnarray}
\label{mond5}
a\(\frac{a}{a_0}\)=G\frac{M}{r^2},
\end{eqnarray}
where 

\begin{eqnarray}
\label{mond6}
a_0=\frac{2\pi c k_B T_c}{\hbar}.
\end{eqnarray}
Using $a_0\approx 10^{-10} m s^{-2}$ we obtain $T_c\approx 10^{-31} K$, an extremely low temperature which is far from the usual temperature observed in our real word.

\section{MOND-Type Theory}

From Eqs.(\ref{mond1}) and (\ref{mond4}) we can deduce that the argument of the interpolation function $\mu(x)$ can be written as $$x\equiv \frac{a}{a_0}=\frac{T}{T_c}\,\,.$$ So, in our MOND-type theory we are considering that the interpolation function is 

\begin{eqnarray}
\label{interpo}
T \leq T_C  \Rightarrow  \mu(x)=\frac{T}{T_c}, \nonumber\\
T > T_C \Rightarrow  \mu(x)=1.
\end{eqnarray}
In Figure 1 we plot $\mu(T)$ as function of $T$. 

\begin{figure}[ih]
\includegraphics[scale=.7,angle=0]{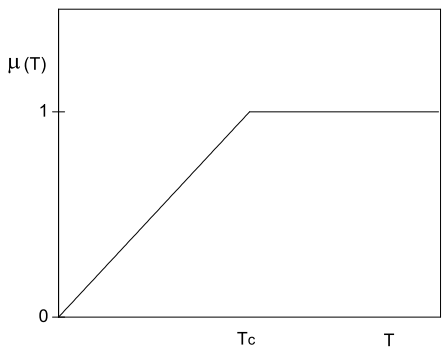}
\caption{An interpolation function curve as function of $T$.} 
\end{figure}

\ni This interpolation function, Eq.(\ref{interpo}), is not a widely-used interpolating function\cite{prof1,prof2} in MOND's theory and because of this reason that we have called our model as MOND-type theory. However, the use of this interpolation function in our approach leads to a second order phase transition phenomenon in MOND's theory where we can derive critical exponents. 

Considering that there is a difference between the number of bits in the holographic screen and the number of bits that exists in the equipartition law, we can choose a particular expression for the $\alpha$-parameter, Eq.(\ref{dif}), that leads to MOND-type theory. To do this we will write Eq.(\ref{mond3}) in the form

\begin{eqnarray}
\label{mond7}
N_S-N_E=N_S\(1-\frac{T}{T_c}\),
\end{eqnarray}
or
\begin{eqnarray}
\label{mond8}
\frac{N_S-N_E}{N_S}=\(1-\frac{T}{T_c}\),
\end{eqnarray}
where $N_S$ and $N_E$ is defined in Eqs.(\ref{dif3}) and (\ref{dif4}) respectively. It is clear that for $T\geq T_c$ we have $\frac{N_S-N_E}{N_S}=0$, since above the critical temperature $T_c$ we have that $N_S=N_E$. Then, from Eq.(\ref{mond7}), the $\alpha$-parameter is given by
 
\begin{eqnarray}
\label{alfa}
\alpha=N_S\(1-\frac{T}{T_c}\),
\end{eqnarray}

\ni which shows that in our MOND-type theory the $\alpha$-parameter depends on the temprature.

\section{Phase transition and critical phenomena analysis}

\ni In order to establish our results within the context of phase transition and critical phenomena, we plot in Figure 2, as an example, a typical spontaneous magnetization curve\cite{Huang}. 
We can observe that for temperatures below $T_c$ the magnetization is different from zero and for temperatures above $T_c$ the magnetization is zero. 

In Figure 3 we have plotted Eq.(\ref{mond8}) as a function of $T$. Also, we can observe that for temperatures below $T_c$ the bits difference number is different from zero and for temperatures above $T_c$ the bits difference number is zero. Comparing both figures we can notice some similarities between the two graphics.

\begin{figure}[ih]
\includegraphics[scale=.7,angle=0]{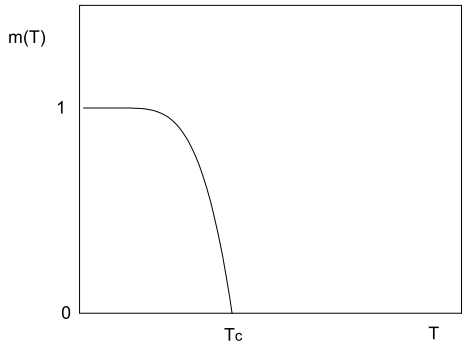}
\caption{A typical spontaneous magnetization curve as function of $T$.} 
\end{figure}

\begin{figure}[ih]
\includegraphics[scale=.7,angle=0]{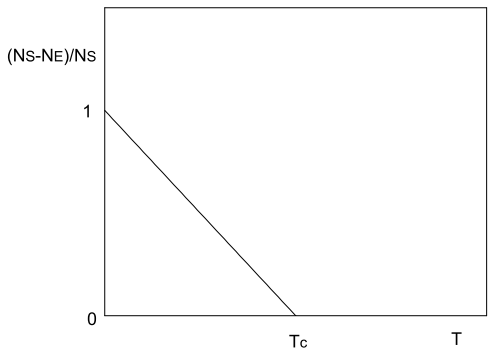}
\caption{The order parameter, $ \frac{N_S-N_E}{N_S}$, as function of $T$.}
\end{figure}

If we consider $\frac{N_S-N_E}{N_S}$ as an order parameter\cite{Huang} then we can observe that the critical exponent $\beta$ associated with the order parameter is equal to 1, i.e.

\begin{eqnarray}
\label{order}
\frac{N_S-N_E}{N_S}\propto\left|\tau\right|^\beta,
\end{eqnarray}
where 
\begin{equation}
\label{AAA}
\tau\equiv 1-\frac{T}{T_c},
\end{equation}

\ni and $\beta=1$. We can use a scaling law for critical exponents, called Josephson scaling\cite{Huang} which can be written as

\begin{eqnarray}
\label{sl}
\nu d=2-\alpha,
\end{eqnarray}
where $\nu$ is the correlation length critical exponent, i.e.

\begin{eqnarray}
\label{len}
\xi(T) \propto\left|\tau\right|^{-\nu},
\end{eqnarray}

\ni and $d$ is the space dimension. In a short description, the correlation length is a measure of the range in which the fluctuations in one region of space are correlated with those in another region. Two points which are separated by a distance larger than the correlation length will each one have fluctuations where they are expected to be independent, i.e., uncorrelated. Experimentally, the correlation length is found to diverge at the critical point.

The derivative of energy $E$ with respect to temperature, which is the heat capacity at constant volume, is zero because in the Verlinde formalism the energy is a constant ($E=M c^2$). Therefore the critical exponent $\alpha$ is zero. 
From the Josephson scaling law, Eq.(\ref{sl}), we obtain that $\nu=2$ since $d=1$ in Verlinde's formalism. So, the correlation length near $T=T_c$ is

\begin{eqnarray}
\label{clen}
\xi(T) \propto\left|\tau\right|^{-2},
\end{eqnarray}

\ni which shows that at $T=T_c \;(\tau=0)$ the correlation length, $\xi(T)$, also diverges in our model. This important result (which confirms what we have mentioned before) indicates that there is a phase transition in the holographic screen from Newton to MOND's regime.

\section{conclusions}

Two recent formalisms that were originated in order to introduce alternative models concerning gravity theory, the so-called MOND and Verlinde's entropic gravity promoted a whole scientific research. The first one had the objective to explain the galaxies' rotation curves and it was considered as an alternative to dark matter.  The second one was aimed to continue the ideas of Beckenstein and Hawking who explored the black holes thermodynamics through entropic concepts and the Unruh thermodynamical acceleration definition, we can obtain Newton's gravitational law.

In this work, we have used both frameworks to obtain an $\alpha$-parameter which has worked as a measure for classical/quantum differences and to analyze the phase transition and critical phenomena.   What is new here about these results is that both are considered on the holographic screen, which is an important concept in the current information theory.

The first result concerning the $\alpha$-parameter has shown that the difference between the number of bits on the holographic screen ($N_S$) and the equipartition law ($N_E$) is not zero, as one would expect.   In other words, since we have shown also that $\alpha$ has quantum features, at the quantum level, $N_S\,-\,N_E$ is not conserved, which could be considered as a kind of anomaly.   The $\alpha$-parameter has also appeared as the perturbative correction of the Newton law and we have compared it with other values obtained in the literature.

Considering MOND, we have obtained another relative value for $N_S\,-\,N_E$.   However, considering phase transition and critical phenomena in the holographic screen, we have demonstrated precisely that the correlation length diverges during the phase transition. Consequently, there is a phase transition in the holographic screen.


\section{Acknowledgments}

\ni EMCA thanks CNPq (Conselho Nacional de Desenvolvimento Cient\' ifico e Tecnol\'ogico), Brazilian scientific support agency, for partial financial support.



\begin{thebibliography}{99}

\bibitem{Verlinde} E. Verlinde, JHEP 1104 (2011) 029.

\bibitem{Pad2} T. Padmanabhan, Mod. Phys. Lett. A, vol. 25, no. 14 (2010) 1129.



\bibitem{MOND1} M. Milgrom, Astrophys. J. 270 (1983) 365.

\bibitem{MOND2} M. Milgrom, Astrophys. J. 270 (1983) 371.

\bibitem{MOND3} M. Milgrom, Astrophys. J. 270 (1983) 384.

\bibitem{unruh} W. G. Unruh, Phys. Rev. D 14 (1976) 870.

\bibitem{Pad1} T. Padmanabhan, arXiv:1206.4916.

\bibitem{Don} J. F. Donoghue, Phys. Rev. D50 (1994) 3874.

\bibitem{Akh} A. Akhundov, S. Bellucci and A. Shiekh, Phys. Lett. B 395 (1997) 16.

\bibitem{BDH} N. E. J. Bjerrum-Bohr, John F. Donoghue and Barry R. Holstein, Phys Rev D 67 (2003) 084033.

\bibitem{TF} R. B. Tully and J. R. Fisher, Astron. Astrophys. 54 (1977) 661.


\bibitem{prof1} B. Famaey, G. Gentile and J. P. Bruneton, Phys. Rev. D 75 (2007) 063002.

\bibitem{prof2} H. S. Zhao and B. Famaey, Astrophys. J. 638 (2006) L9.

\bibitem{JAN} Jorge Ananias Neto, Int. J. Theor. Phys. 50 (2011) 3552.


\bibitem{Ep1} E. Pazy, Phys. Rev. D  85 (2012) 104021.

\bibitem{Ep2} E. Pazy, Phys. Rev. D  87 (2013) 084063.

\bibitem{Huang} K. Huang, Statistical Mechanics, 2nd edn. Wiley, (1987) New York.

















\end{thebibliography}
\end{document}